\documentclass[aps,pre,reprint]{revtex4-1}

\draft 

\usepackage{graphicx}
\usepackage{dcolumn}
\usepackage{bm}
\usepackage{color}
\usepackage{float}
\usepackage{hyperref}
\usepackage[version=4]{mhchem}
\usepackage{relsize}
\usepackage{soul}

\begin{document}

\title{Excitation of Faraday-like body waves in vibrated living earthworms}

\author{Ivan S. Maksymov}
\affiliation{Centre for Micro-Photonics, Swinburne University of Technology, Hawthorn, Victoria, 3122, Australia\looseness=-1}
\email{imaksymov@swin.edu.au}
\author{Andrey Pototsky}
\affiliation{Department of Mathematics, Faculty of Science Engineering and Technology, Swinburne University of Technology, Hawthorn, Victoria, 3122, Australia}
\email{apototskyy@swin.edu.au}


\begin{abstract}
Biological cells and many living organisms are mostly made of liquids and therefore, by analogy with liquid drops, they should exhibit a range of fundamental nonlinear phenomena such as the onset of standing surface waves. Here, we test four common species of earthworm to demonstrate that vertical vibration of living worms lying horizontally of a flat solid surface results in the onset of subharmonic Faraday-like body waves, which is possible because earthworms have a hydrostatic skeleton with a flexible skin and a liquid-filled body cavity. Our findings are supported by theoretical analysis based on a model of parametrically excited vibrations in liquid-filled elastic cylinders using material parameters of the worm's body reported in the literature. The ability to excite nonlinear subharmonic body waves in a living organism could be used to probe, and potentially to control, important biophysical processes such as the propagation of nerve impulses, thereby opening up avenues for addressing biological questions of fundamental impact. 

\end{abstract}

\maketitle 

\section{Introduction}

Sound and vibrations are essential for efficient communication between living beings and they also underpin human-made imaging, spectroscopy and sensing techniques such as medical ultrasound and photoacoustic imaging modalities \cite{Eme19}, Brillouin Light Scattering spectroscopy \cite{Bal19}, and laser vibrometry \cite{Bla13} to name a few. Sound and vibrations are also likely to play an important role in the propagation of nerve impulses \cite{Hei05, Had15} as well as they can be used to develop new methods of bacteria and virus killing \cite{Zin05, Iva13, boyd2019beamed}. Furthermore, using vibrations one could monitor, understand and control the behaviour of some animals, such as earthworms \cite{Mit08}, and exploit them to sense and modify soil structure as well as to increase crop yields \cite{Bla17, Lac18, Rui18}.

Earthworms -- tube-shaped, segmented worms that have a world-wide distribution and are commonly found in soil -- have become a subject of intensive research focused on their response to vibrations and sound. Some of these studies aim to explain the response of these animals to natural vibrations produced by predators, rain or plants \cite{Mit08}. Furthermore, the glial cell wrapping of the giant axons of earthworms resembles the myelin sheath of vertebrate nerve fibers \cite{Roo83}. Therefore, earthworms serve as a platform for neurobiological studies \cite{Sha14}. Earthworms are also cheap and using them does not require ethics approval. Hence, we choose these animals to demonstrate the onset of Faraday-like subharmonic body waves in a living organism subjected to external mechanical vibration.

Classical nonlinear standing Faraday waves appear on the horizontal surface of an infinitely extended liquid supported by a vertically vibrating container \cite{Faraday}. For any given vibration frequency $\omega$, when the vibration amplitude exceeds a certain critical value, the flat surface of the fluid becomes unstable and subharmonic surface waves oscillating at the frequency $\omega/2$ are formed. These oscillations are due to a parametric resonance between the forcing at the frequency $\omega$ and gravity-capillary surface waves with the dispersion relation $\Omega(k)$, being $k$ a certain wave vector selected as $\Omega(k)=\omega/2$. 

Faraday waves have become a paradigmatic example of nonlinear wave systems exhibiting complex periodic \cite{Benjamin_54} and quasi-periodic \cite{henderson_miles_1990,miles_1984,jiang_1996} dynamics as well as chaotic behaviour \cite{Punzmann09,Xia12,Shats2010,Shats2012}. Recently, a number of applications of Faraday waves in the fields outside the area fluid dynamics have been suggested, including novel photonic devices \cite{Tarasov_2016, Huang_17}, metamaterials \cite{Domino_2016, Fra17}, alternative sources of energy \cite{Alazemi_2017}, and applications in biology \cite{Sheldrake_2017}.

Parametrically excited vibrations and surface waves have also been observed in isolated liquid drops subjected to external mechanical forcing \cite{Yoshiyasu96,Noblin09,Ma17,ma_burton_2018,Hemmerle_2015, Pucci_2011,Pucci_2013, Pototsky_2018,maksymov2019harmonic}. In response to vibration, the drop can either adopt a regular star shape \cite{Yoshiyasu96,Noblin09,Ma17,ma_burton_2018} or exhibit a more dramatic transformation by spontaneously elongating in horizontal direction to form a worm-like structure of gradually increasing length \cite{Pucci_2011,Pucci_2013,Hemmerle_2015, Pototsky_2018}.

In contrast to the classical Faraday instability in infinitely extended systems, in isolated liquid drops the boundary conditions at the drop edge dictate the existence of a discrete set of vibrational modes \cite{Bostwick1,Bostwick2,Bostwick3,Chang_2013,Chang_2015}. The eigenfrequency $\Omega$ of each mode depends on the boundary conditions at the contact line \cite{Bostwick2,Bostwick3}. When a drop is vibrated at the frequency $\omega$, the fundamental subharmonic resonance occurs when the resonance condition $\Omega=\omega/2$ is fulfilled \cite{Yoshiyasu96}.

In inviscid fluids, the subharmonic response sets in at a vanishingly small vibration amplitude at frequencies that satisfy the resonance condition. For frequencies that do not satisfy the resonance condition, the critical amplitude is nonzero. In experiments with viscous isolated drops, the dependence of the subharmonic critical amplitude on the vibration frequency $\omega$ was shown to exhibit periodic variations \cite{Yoshiyasu96,Ma17,ma_burton_2018}. This feature is in stark contrast with the Faraday instability in infinitely extended fluids, where the critical amplitude monotonically increases with the driving frequency $\omega$ \cite{Kumar96}.

In this work, we observe experimentally the subharmonic oscillations of the body of living earthworms lying horizontally on a flat solid surface subjected to vertical vibration. We measure the critical amplitude of the onset of subharmonic response as a function of the vibration frequency $f$, and we reveal that the obtained dependence exhibits signature characteristics of parametrically excited capillary surface waves in vibrated liquid drops \cite{Yoshiyasu96,Noblin09,Ma17,ma_burton_2018}. In particular, we show that the critical amplitude varies periodically with $f$. We explain the observed results by modelling the body of the worm as a horizontally-extended, liquid-filled elastic cylinder subjected to vertical vibration. 

Because the excitation of Faraday-like waves in living organisms has thus far received little attention \cite{hong2019wavecontrolled}, our findings promise to push the frontiers of our knowledge of fundamental nonlinear phenomena and chaotic behaviour in biological systems. For instance, our results should be qualitatively reproducible in other living systems such as bacteria, biological cells or individual organs in the body including the brain and blood vessels. 

\section{Experimental results}

We tested four different earthworm species encountered in the south eastern regions of Australia \cite{Bak06}. To correctly identify the earthworm species, we used an earthworm identification guide \cite{OPAL}. \textit{Eisenia fetida} earthworms were purchased from a local fishing goods store, and on average they were $100-120$\,mm long and $5-6$\,mm wide. \textit{Lumbricus terrestris} earthworms were harvested in the field and closely related to them \textit{Lumbricus rubellus} earthworms were obtained from a local compost worm supplier. In this group, we selected the worms that measured approximately $120-150$\,mm in length and $8-10$\,mm in width. Several smaller $6-8$-mm-long and $2-3$-mm-wide \textit{Aporrectodea caliginosa} earthworms were also harvested in the field and outcomes of their test were qualitatively similar. 

Earthworms are non-regulated animals, and therefore this research did not require the approval of our Institutional Animal Ethics Committee. However, the worms were treated as humane as practical and afterwards they were placed into a worm farm where they fully recovered. 

\begin{figure}[t]
\centering
\includegraphics[width=0.475\textwidth]{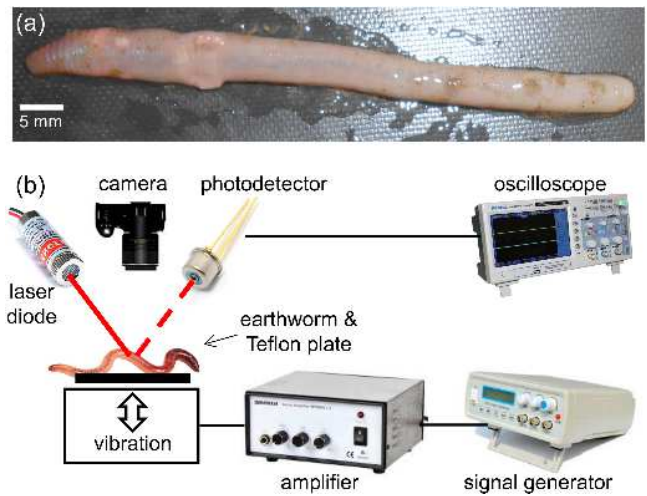}
\caption{(a) Photograph of an anaesthetised \textit{Eisenia fetida} earthworm. (b) Schematic of the experimental setup. A subwoofer covered by a thin Teflon plate is used as the source of vertical vibration. The sinusoidal vibration signal of frequency $f$ is synthesised with a digital signal generator and amplified with an audio amplifier. Vibrations of the earthworm placed horizontally on top of the Teflon plate are measured by using a continuous wave red laser diode and a photodetector. The detected signals are visualised with an oscilloscope and sent to a laptop for post-processing. A digital camera is used to continuously monitor the position of the worm.
\label{Fig1}}
\end{figure}

In preparation for experiments, earthworms were first placed in $20\%$ ethanol for approximately $2$ minutes, which immobilised them to simplify handling. Then, the entire body of an immobilised worm was placed on top of a thin Teflon plate [Fig.~\ref{Fig1}(a)] that was vertically vibrated with the harmonic frequency $f$. The vibrations were detected by using an in-house laser vibrometry setup\cite{maksymov2019harmonic} [Fig.~\ref{Fig1}(b)] consisting of a red laser diode (Besram Technology, China, $650$\,nm wavelength and $1$\,mW maximum power) and a photodetector (Adafruit, USA). The intensity of light reflected from the worm is modulated due to the vertical vibration as well as the onset of parametrically excited body waves. We recorded these signals with Audacity software and Fourier-transformed them with Octave software to obtain frequency spectra. The skin of the worm was moistened with water to avoid drying during the study. Excessive liquid was removed from the Teflon plate to ensure that Faraday waves are not excited on the liquid surface. 

The experiments were conducted by using the following protocol. The laser beam was focused on the body of the worm and the vibration amplitude of the Teflon plate was gradually increased until the point when the onset of Faraday instability was observed with the oscilloscope. The position of the worm was continually monitored with a digital camera to make sure that the same part of the worm is illuminated. Large vibration amplitudes leading to a horizontal displacement of the worm or jumps of the entire body were avoided to help keep the worm in the focus of the laser vibrometry setup. Large vibrations were also avoided because they additionally lead to ejection of a sticky fluid from the worm, which could serve as a medium for surface Faraday waves not wanted in our measurements. Typically, the total duration of measurements using the same worm was under five minutes to avoid desiccation. After the experiment, the worms were rehydrated and released into a worm farm. 
  
\begin{figure}[t]
\centering
\includegraphics[width=0.475\textwidth]{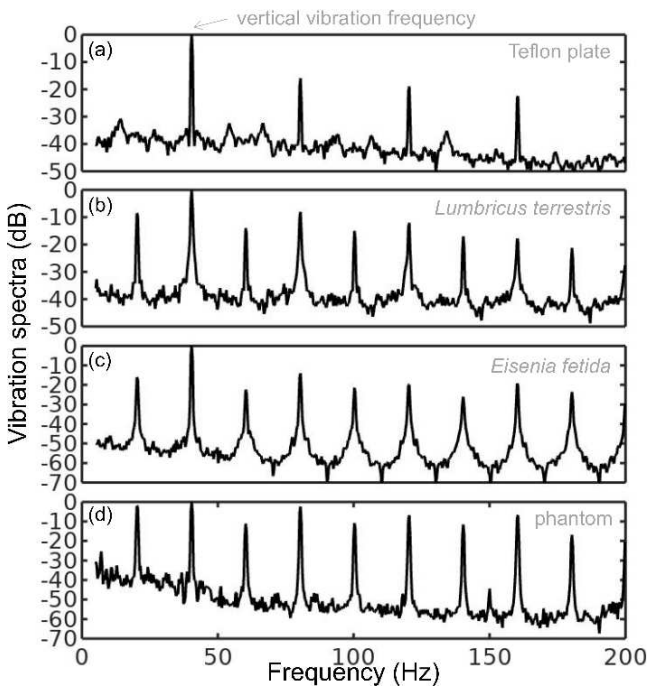}
\caption{Vibration spectra of the (a) Teflon plate without the worm, (b) \textit{Lumbricus terrestris} earthworm, (c) \textit{Eisenia fetida} earthworm and (d) earthworm-mimicking phantom. The vertical vibration frequency is $f=40$\,Hz. All spectra are normalised to their respective maxima. Note the presence of subharmonic ($f/2, 3f/2$ and so on) frequency peaks in the spectrum of the earthworms and the phantom, which are not present in the spectrum of the unloaded Teflon plate. The peaks in the spectrum of the earthworms are wider than in the case of the Teflon plate, which is a result of the amplitude modulation and appearance of frequency sidebands leading to the broadening of the peak \cite{maksymov2019harmonic}. Also note that the spectra of the worms and the worm-mimicking phantom are qualitatively similar.
\label{Fig2}}
\end{figure}

Figures~\ref{Fig2}(a) and (b) show, respectively, the vibration spectrum of the Teflon plate without the worm and the vibration spectrum of a \textit{Lumbricus terrestris} earthworm placed horizontally on top of the Teflon plate. In both cases the Teflon plate is subjected to vertical vibration at $f = 40$\,Hz. 

The vibration spectrum of the unloaded Teflon plate is dominated by the peak at the frequency $f = 40$\,Hz and its higher-order harmonic frequencies $80$, $120$ and $160$\,Hz. The intensity of the second (third) harmonic is approximately $50$ ($80$) times smaller than that of the fundamental signal and these signals are consistent with the intrinsic nonlinear distortion of the source of vertical vibrations used in our setup.  

In the vibration spectrum of the earthworm, we observe that the harmonic waves lose their stability via a period-doubling bifurcation \cite{maksymov2019harmonic} as evidenced by the appearance of the peaks at $f = 20, 60, 140$ and $180$\,Hz not present in the vibration spectrum of the Teflon plate. Significantly, the harmonic frequency peaks in the spectrum of the worm are wider as compared with the respective peaks in the spectrum of the Teflon plate. This is because the onset of subharmonic oscillations also results in the amplitude modulation and appearance of sidebands, which in turn leads to broadening of the spectral peaks previously reported for liquid films \cite{Punzmann09,Xia12} and liquid drops subjected to vibration \cite{maksymov2019harmonic}. Qualitatively the same behaviour is observed in the vibration spectrum of the other tested earthworm species, including the \textit{Eisenia fetida} worms [Fig.~\ref{Fig2}(c)]. 

The analogy between the worm and liquid drop is possible because earthworms have a hydrostatic, supported by fluid pressure skeleton with a flexible skin and a liquid-filled body cavity. When the worm is anaesthetised, its nervous system does not produce nerve impulses and therefore the muscles of its body are fully relaxed \cite{Sha14}. Moreover, whereas in non-anaesthetised worms high internal pressure helps to maintain a cylindrical shape \cite{Har57}, the pressure in anaesthetised worms is low \cite{McK87}. We verified this by establishing that a puncture of the skin of an anaesthetised worm does not lead to a dramatic ejection of internal fluids typically seen in non-anaesthetised worms. (Here, we draw an analogy between the worm and a pressurised balloon.) Hence, it is plausible to consider the worm to be a liquid drop enclosed by a thin elastic skin.

To verify this hypothesis, we test an earthworm-mimicking phantom made of a finger of an approximately $0.1$-mm-thick latex glove filled with water. The thickness of the phantom wall is of the same order of magnitude as the body wall of an earthworm \cite{BRIONES201826}, but mechanical properties of latex films of sub-millimeter thickness \cite{Sim12} are similar to those of living worms \cite{Backholm_2013}. The vibration spectrum of the phantom is shown in Fig.~\ref{Fig2}(d) and it is in good qualitative agreement with the vibration spectra of the real earthworms.

\begin{figure}[t]
\centering
\includegraphics[width=0.35\textwidth]{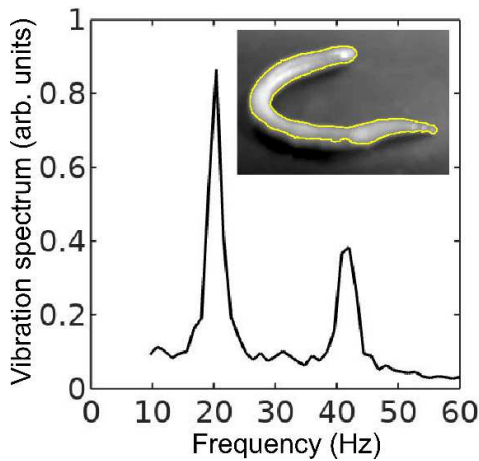}
\caption{Vibration spectrum of the body of the earthworm subjected to vertical vibration at the frequency $f=40$\,Hz obtained by processing a video of the vibrated worm as explained in the main text. The subharmonic peak at $f/2=20$\,Hz can be clearly seen. The inset shows an example of the contour of the worm obtained from a single frame extracted from the video.
\label{Fig_body_vibration}}
\end{figure}

Furthermore, in the following we assume that the body wall of the worm is an elastic cylindrical shell undergoing flexural vibrations. (A relevant model was used in Ref.~\cite{Rui18}, but in Sec.~\ref{model} of this paper we show that flexural vibrations should be responsible for the subharmonic response of the worms.) Direct mapping of flexural vibration modes with our laser vibrometry setup is currently unavailable. Thus, to demonstrate that the entire body of worm undergoes vibrations resulting in the appearance of subharmonic frequencies in the vibration spectrum, we use a time-domain approach. We first film worms  vibrated at $40$\,Hz frequency with a digital camera capturing $120$ frames per second and then we process the resulting videos in Octave software, where we binarise each frame and use the \texttt{bwboundaries} command to find the contours of the worm (see the inset in Fig.~\ref{Fig_body_vibration}). Then, we calculate the area of the contour for every frame and we finally Fourier-transform the resulting area-versus-time dependence to obtain the vibration spectrum of the body of the worm.

As shown in Fig.~\ref{Fig_body_vibration}, in agreement with the results obtained with the laser vibrometry setup (Fig.~\ref{Fig2}), the body of the worm oscillates at the vibration frequency $f \approx 40$\,Hz and the subharmonic frequency $f/2 = 20$\,Hz. However, whereas the spectra in Fig.~\ref{Fig2} were obtained by focusing the laser beam on a part of the worm's body, the spectrum in Fig.~\ref{Fig_body_vibration} originates from the vibrations of the entire body of the worm. The higher frequencies peaks present in Fig.~\ref{Fig2} are not reproduced in Fig.~\ref{Fig_body_vibration}, because the finite resolution and frames-per-second speed of the digital camera limit the resolving ability of the image processing based approach at higher frequencies. 

\begin{figure}[t]
\centering
\includegraphics[width=0.35\textwidth]{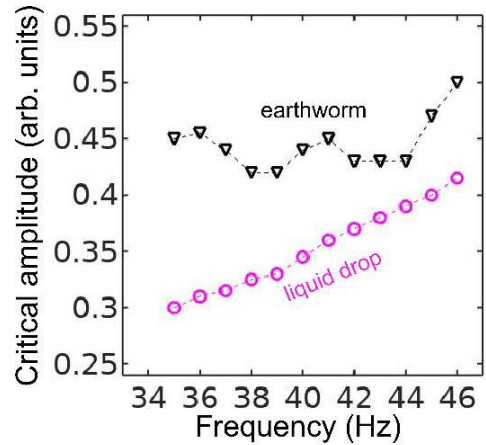}
\caption{Critical vibration amplitude of the onset of the subharmonic response in an earthworm (triangles) and a pancake-like canola oil drop (circles) plotted as a function of the vertical vibration frequency $f$. The dashed lines are the guide to the eye. Whereas in the selected frequency range the response of the canola oil drop is quasi-monotonic, the curve for the critical amplitude for the worm exhibits oscillations. These oscillations allows us to correlate the experimental data with the predictions of our theoretical model developed in Sec.\,\ref{model}.
\label{Fig3}}
\end{figure}

As a next step, we measure the lowest value of the vibration amplitude at which the subharmonic response of the worm body sets in. We call this value the critical amplitude and we plot it in Fig.~\ref{Fig3} as a function of the frequency $f$ for an \textit{Eisenia fetida} worm.

In the $35...45$\,Hz frequency range, the critical amplitude is relatively low and therefore the body of the worm does not shift along the surface of the Teflon plate, thereby allowing us to obtain accurate results. In the critical amplitude dependence for the worm (Fig.~\ref{Fig3}), we observe oscillations with the two minima at $f\approx38$\,Hz and $f\approx43$\,Hz. In contrast, the critical amplitude of an approximately $6$\,cm$\times 6$\,cm pancake-like drop of canola oil is quasi-monotonic in the $35...45$\,Hz range. [However, in agreement with the previous results \cite{Yoshiyasu96,Ma17,ma_burton_2018}, we observed a non-monotonic response at $f>50$\,Hz (not shown for simplicity)].

A similar nonmonotonic dependence of the critical amplitude on the vibration frequency was previously observed in infinitely extended viscoelastic films \cite{Bal05}. In contrast to liquid drops of simple Newtonian fluids \cite{maksymov2019harmonic}, the nonmonotonic dependence originates from the ability of a viscoelastic material to 'remember' past stresses. Thus, in the Maxwell model of linear elasticity, the instantaneous stress in the material is described by a time-dependent relaxation modulus decaying over a characteristic relaxation time. The coupling between the period of forcing with the relaxation time of the material viscoelastic response leads to the oscillation of the critical amplitude as the function of the vibration frequency.

Based on our experimental data, it is not possible to conclude whether the nonmonotonic dependence in Fig.~\ref{Fig3} is due to viscoelastic properties of the worm tissues, the finite size of the worm body or a combined effect of viscoelastic properties and the geometry of the worm. In the following, we assume that, similar to small liquid drops \cite{Yoshiyasu96,Noblin09,Ma17,ma_burton_2018}, the nonmonotonic subharmonic response of the worm originates from the discrete spectrum of its natural vibration frequencies. In Sec.~\ref{model} we develop a theoretical model considering the worm as an elastic cylindrical shell filled with an incompressible fluid, and we correlate the experimental results with the theoretical predictions.

\section{Theoretical model}
\label{model}
The observed subharmonic oscillations of the earthworms bear a striking resemblance to the well-known phenomenon of parametrically excited capillary surface waves in vertically vibrated liquid drops \cite{Yoshiyasu96,Noblin09,Ma17,ma_burton_2018}. 

\begin{figure}[t]
\centering
\includegraphics[width=0.5\textwidth]{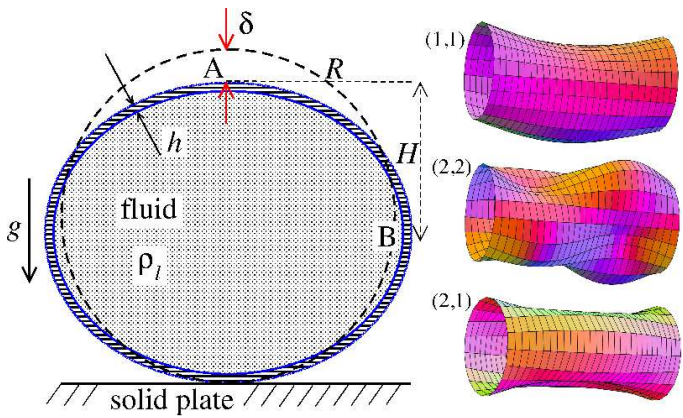}
\caption{ Schematic of the theoretical model for a vibrated earthworm represented as a liquid-filled elastic cylinder of length $L$ with the Young's modulus $E$, undeformed radius $R$, and shell thickness $h$. The cylinder is filled with incompressible and inviscid fluid with the density $\rho_l$. (left) Cross-section of the elastic cylinder deformed by the gravity in contact with a non-deformable solid plate. The dashed line shows the contour of an undeformed cylinder of radius $R$. The squashing depth is $\delta\approx R\sqrt{\rho g R^2/(3Eh)}$. (right) Calculated spatial profiles of the first three lowest frequency vibration modes. The integer numbers in the parentheses denote, respectively, the number of the circumferential and the axial vibration mode (see the main text). }
\label{Fig5}
\end{figure}

To understand the physical origin of the subharmonic response of earthworms, we neglect the damping effect of the viscosity and model the body wall of the worm as an elastic cylindrical shell of length $L$ with the Young's modulus $E$, radius $R$ and shell thickness $h$. The cylinder is filled with incompressible and inviscid fluid with the density $\rho_l$. 

An elastic cylindrical shell model of the worm body has been previously used to analyse the pressure exerted by earthworms during their burrowing activity \cite{Rui18}. When such a cylinder with the horizontally oriented axis is supported by a solid plate, its cross-sectional shape is no longer circular due gravity deformation [Fig.\,\ref{Fig5}(left)]. To estimate the squashing depth $\delta$ that measures the change in the vertical height of the cylinder in the squashed state, we neglect the bending energy of the thin shell ($h\ll R$) as compared with the energy due to stretching. In this case, the equilibrium shape of the gravity deformed cross-section of the cylinder filled with an incompressible fluid of density $\rho_l$ can be found by balancing the tension $T$ per unit axial length of the elastic shell with the hydrostatic pressure \cite{Wang81}. In particular, we obtain $\rho_l g H=T(\kappa_B-\kappa_A)$, where $ H$ is the height difference between the points $A$ and $B$ in Fig.\ref{Fig5}(left) and $\kappa_{A,B}$ denotes the curvature of the shell at the points $(A,B)$. Assuming weak deformation such that $\delta/R \ll 1$, we approximate the shape of the squashed cylinder by an ellipse with the minor and the major semi-axis $R-\delta/2$ and $R+\delta/2$, respectively. Then, the curvatures to the first order in $\delta/R$ are $\kappa_A=(R-\delta/2)/(R+\delta/2)^2\approx R^{-1}-3\delta/(2R^2)+ O(\delta/R)^2$ and $\kappa_B=(R+\delta/2)/(R-\delta/2)^2\approx R^{-1}+3\delta/(2R^2)+O(\delta/R)^2$.  Finally, taking into account that $T=Eh\delta /R$ and $H= R-\delta/2$, we obtain the following estimate
\begin{eqnarray}
    \label{mod_eq0}
    \frac{\delta}{R}= \sqrt{\frac{\rho g R^2}{3Eh}}.
    \end{eqnarray}

Viscoelastic properties of earthworms are poorly understood. However, mechanical properties of millimeter-sized nematode \textit{Caenorhabditis elegans} worms have been recently reported \cite{Backholm_2013}. In particular, the effective bulk Young's modulus of the worm body was found to be in the $10^5...10^6$\,Pa range, with the maximum value originating from the cuticle. On the other hand, it has also been suggested that all nematodes exhibit a universal elastic response dominated by the mechanics of pressurised internal organs \cite{Gil15}. Nevertheless, the values of the bulk modulus reported in \cite{Gil15} are of the same order of magnitude as those in \cite{Backholm_2013}. Furthermore, the effective Young's modulus of the cuticle strongly depends on its thickness \cite{Backholm_2013, Gil15}. Thus, assuming that the elasticity of the worm body is entirely due to the stiffness of the cuticle, the Young's modulus can reach $200...400$\,MPa \cite{Nakajima09}.

Significantly, for the lowest expected value of the effective Young's modulus of the cuticle $E=1$\,MPa \cite{Backholm_2013, Gil15} the gravity squashing of the worm body remains small, as demonstrated below. We take the thickness of the cylindrical wall to be $h=50$ $\mu$m, which corresponds to the combined thickness of the cuticle and the epidermis of earthworms \cite{BRIONES201826}. The density of the internal body fluid can be estimated as $\rho_l=1100$\,kg$/$m$^3$. This estimate is warranted because, for example, the worm has blood vessels filled with blood, but the density of blood plasma is approximately $1025$ kg$/$m$^3$ and the density of blood cells circulating in the blood is approximately $1125$\,kg$/$m$^3$. Finally, for the largest possible value of the worm radius used in our experiments $R=5$\,mm, we estimate the relative squashing from Eq.\,(\ref{mod_eq0}) to be $(\delta/R)\approx 4\%$. We note that in reality $(\delta/R)$ should be much smaller than $4\%$ because the effective Young's modulus of the cuticle could be about three orders of magnitude higher \cite{Nakajima09}.

Following the approach developed for liquid drops \cite{Yoshiyasu96}, we consider the natural vibrational eigenfrequencies of a gravity squashed cylinder [Fig.\,\ref{Fig5}(left)]. To the best of our knowledge, the analysis of the vibrational frequencies of a cylinder supported by a solid plate along its entire length has not been reported. However, one can estimate the order of magnitude of the natural frequencies by using the well-known result for a cylinder with freely supported ends \cite{Arnold49,Warburton65,Lindholm62}, which is demonstrated below.

In cylindrical coordinates, the axial $u(r,\theta)$, azimuthal $v(r,\theta)$ and radial $w(r,\theta)$ displacements of the shell of a cylinder with freely supported ends can be written as
\begin{eqnarray}
\label{mod_eq1}
u&=&U(t)\cos{(\pi m z/L)}\cos{(n\theta)},\\
v&=&V(t)\sin{(\pi m z/L)}\sin{(n\theta)},\\
w&=&W(t)\sin{(\pi m z/L)}\cos{(n\theta)},
\end{eqnarray}
where the integers $n$ and $m$ determine the circumferential and the axial vibration modes, respectively. As a representative example, in Fig.~\ref{Fig5}(right) we show the first three lowest frequency modes.

For $\delta/R \ll 1$, the deviation of vibrational frequencies from those of a circular cylinder is of first order in $\delta/R$ \cite{SHIRAKAWA82}. This allows us to write the equations of motion for the amplitudes $U,V,W$ as 
\begin{eqnarray}
\label{mod_eq2}
\ddot{{\bm x}}\approx \left({\bm J}_0+\frac{\delta}{R}{\bm J}_1\right){\bm x},
\end{eqnarray}
where ${\bm x}=(U,V,W)$, ${\bm J}_0$ is the Jacobi matrix corresponding to the circular cylinder and ${\bm J}_1$ is its first order correction due to squashing. When the solid plate is vibrated with the frequency $\omega$, its vertical displacement is given by $A\cos{(\omega t)}$, where $A$ is the vibrational amplitude. The gravity acceleration in the co-moving frame of reference is $g(t)=g[1+a\cos{(\omega t)}]$, where $a=A\omega^2/g$ is the dimensionless scaled amplitude. Since viscosity is neglected, we anticipate the onset of the subharmonic vibrations at small amplitude $a \ll 1$. In this regime, we obtain from Eq.\,(\ref{mod_eq0}) for the time-dependent squashing $\delta/R\approx  \sqrt{\frac{\rho g R^2}{3Eh}}\left(1+\frac{a}{2}\cos{(\omega t)}\right)$. In this limit, Eq.\,(\ref{mod_eq2}) reduces to the three-dimensional Mathieu equation
\begin{eqnarray}
\label{mod_eq3}
\ddot{{\bm x}}= \left({\bm J}_0+\sqrt{\frac{\rho g R^2}{3Eh}}{\bm J}_1+q\cos{(2\pi f)}{\bm J}_1\right){\bm x},
\end{eqnarray}
where $q=\frac{a}{2}\sqrt{\frac{\rho g R^2}{3Eh}}$ is the scaled vibration amplitude.

The properties of solutions of Eq.\,(\ref{mod_eq3}) are well-known \cite{hsu1963parametric}. Parametrically excited instability sets in when $2\pi f$ coincides with one of the combination frequencies $|(\omega_0)_i\pm (\omega_0)_k|$, where $(\omega_0)_i^2$ $(i=1,2,3)$ are the eigenvalues of ${\bm J}_0+\sqrt{\frac{\rho g R^2}{3Eh}}{\bm J}_1$. Because $\sqrt{\frac{\rho g R^2}{3Eh}}\sim 0.04$ for $E=1$\,MPa, we can neglect the term $\sqrt{\frac{\rho g R^2}{3Eh}}{\bm J}_1$ in Eq.\,(\ref{mod_eq3}) as compared with ${\bm J}_0$, which implies that the natural frequencies $(\omega_0)_i$ of the gravity squashed worm can be approximated by those of a circular elastic cylinder. 

Amongst the three frequencies $(\omega_0)_i$ $(i=1,2,3)$ one is typically two orders of magnitude lower than the other two. This lowest frequency $\omega_0$ corresponds to the mode with predominantly radial displacement and is given by \cite{Lindholm62} 

\begin{eqnarray}
\label{freq}
\frac{\omega_0^2 R^2\rho}{E}&=&\left(\frac{h^2(\lambda_m^2+n^2)^2}{12R^2(1-\nu^2)}+\frac{\lambda_m^4}{(\lambda_m^2+n^2)^2}\right)\times \nonumber \\
&&\left(1+\frac{a\rho_l I_n(\lambda_m)}{\lambda_m h\rho (I_n)'(\lambda_m)}\right)^{-1},
\end{eqnarray}
where $\lambda_m=m\pi R/L$ and $I_n$ is the modified Bessel function of the first kind of order $n$.

\begin{figure*}[t]
\centering
\includegraphics[width=0.75\textwidth]{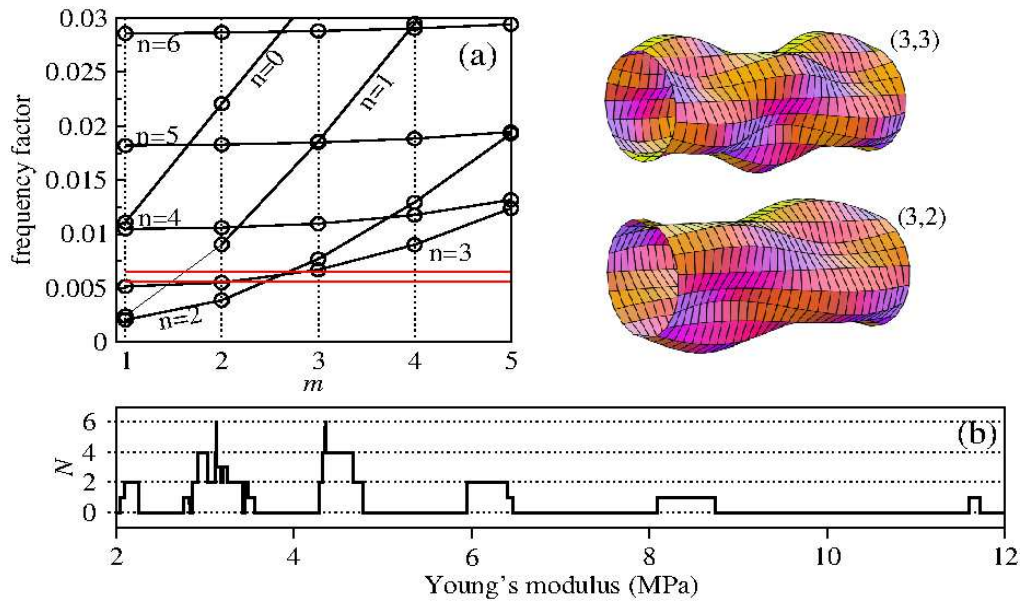}
\caption{Analysis of experimental results in light of the predictions of the developed theoretical model. By scanning through all theoretically possible combinations of vibration modes and varying the value of the effective Young's modulus of the worm, we find the modes involved in the subharmonic response of the worm and plot the respective spatial mode profiles. (a) Frequency factor $\omega_0 R \sqrt{\rho/E}$ plotted as a function of the axial mode number $m$ for the circular elastic cylinder filled with liquid with density $\rho_l$. The other parameters are $R=5$\,mm, $L=10$\,cm, $\rho=\rho_l=1100$\,kg$/$m$^3$, $\nu=0.5$ and $h=50$\,$\mu$m. The circumferential mode number $n$ is indicated next to each curve. The two solid horizontal lines correspond to the levels of $38\pi R \sqrt{\rho/E}$ and $43\pi R \sqrt{\rho/E}$ with $E=8.3$\,MPa. For $E=8.3$\,MPa, the mode $(n=3,m=3)$ is excited at $38$\,Hz and the mode $(n=2,m=3)$ is excited at $43$\,Hz. The frequencies $38$\,Hz and $43$\,Hz correspond to the first and the second minimum of the critical vibration amplitude function in Fig.~\ref{Fig3}. (b) The number of modes $N$ that match the subharmonic resonance criterion as a function of $E$ (see the main text for details). From this panel we obtain information about the largest possible values of the Young's modulus corresponding to the modes at $38$\,Hz and $43$\,Hz.
\label{Fig6}}
\end{figure*}

Considered as a whole, the combination $\omega_0 R\sqrt{\frac{\rho}{E}}$ in Eq.~\ref{freq} (called the frequency factor) depends on the geometry of the cylinder, the Poisson ratio $\nu$ and the ratio of the densities $\rho_l/\rho$, but is independent of the Young's modulus $E$. In Fig.~\ref{Fig6}(a), we plot the frequency factor as a function of the axial mode number $m$ for different values of $n$ for a cylinder of length $L=10$\,cm, shell thickness $h$=50\,$\mu$m, radius $R=5$\,mm, filled with a fluid with density $\rho_l=\rho=1100$\,kg/m$^3$. The first three lowest frequency modes with $(n=1,m=1)$, $(n=2,m=2)$ and $(n=2,m=1)$ are shown in the right panel of Fig.~\ref{Fig5} for the sake of illustration.

\section{Discussion}

The theory developed in Sec.~\ref{model} can, in general, be used to find the mode profiles corresponding to the resonance frequencies found in the experiment in Fig.~\ref{Fig3} by using experimental data for the Young's modulus $E$ of earthworms as a key input parameter. However, plausible values of $E$ for different worm species are a subject of active debate due to a large range of the reported values and poor understanding of the impact of the cuticle on mechanical properties of worms \cite{Backholm_2013, Gil15}.

To circumvent the lack of experimental data, we reanalyse our experimental results in Fig.\ref{Fig3} in light of the predictions of the developed model Eqs.\,(\ref{mod_eq3},\ref{freq}). In particular, we establish which of the theoretically possible vibrational modes could be excited at a given value of $E$. Here, we vary the value of $E$ in a range bounded by two critical values -- the effective bulk Young's modulus of the worm from Ref.~\cite{Backholm_2013} and the locally measured stiffness of the cuticle \cite{Nakajima09, Gil15}. Whereas the exact values of $E$ are yet to be confirmed experimentally, it has already been established that $E$ would be a function of the thickness of the cuticle \cite{Gil15} and that it would approach $200...400$\,MPa \cite{Nakajima09} in a limiting case of the mechanical properties of the worm defined solely by the cuticle.

Naturally, the thickness and stiffness of the cuticle vary for different species of worms and they are also likely to vary from one animal to another within the same species group. Indeed, in our experiments we established that \textit{Eisenia fetida} earthworms appear to be slightly stiffer when palpated as compared with the other species tested in this work. However, this difference alone cannot result in an order of magnitude discrepancy in the values of $E$.

Firstly, for the fundamental subharmonic resonance in our model we choose the doubled frequency to be $\omega_0=\pi f_{\rm min}$, where $f_{\rm min}$ is either $38$\,Hz or $43$\,Hz corresponding to the first and the second minimum of the measured critical vibration amplitude function (Fig.~\ref{Fig3}). Then, for any fixed value of $E$ we find all theoretically possible vibrational modes $(n=0,\dots,m=1,\dots)$ whose doubled frequency $\omega_0/\pi$ differs by at most $1$\,Hz from either $38$\,Hz or $43$\,Hz. We choose the tolerance of $\pm 1$ Hz because it corresponds to the frequency resolution in the experimental data in Fig.\ref{Fig3}. It is noteworthy that our model would predicts another upper bound for $E$ when the tolerance is varied. However, choosing the tolerance dictated by the resolution in the experimental data serves the purpose of comparing the experimental and theoretical results obtained in this work.

Denoting the number of modes that match $38$\,Hz and $43$\,Hz as $N_{38}$ and $N_{43}$, respectively, we define the total number of matches as $N=N_{38}N_{43}$. Finally, by gradually varying $E$ with a fixed increment, we obtain the dependence $N(E)$ and plot it in Fig.\,\ref{Fig6}(b). We observe that the largest possible value of the Young's modulus lies in the $E=8.3...8.9$\,MPa range, which corresponds to the mode $(n=3,m=3)$ excited at $f=38$\,Hz and the mode $(n=2,m=3)$ excited at $f=43$\,Hz. The corresponding values of the frequency factor are shown by the two horizontal lines in Fig.\,\ref{Fig6}(a). Based on these results, we predict that the body of the worm subjected to vertical vibration and undergoing Faraday-like body oscillations would assume the spatial profiles shown in the insets in Fig.\,\ref{Fig6}(a).

The values of the Young's modulus $E{=}8.3...8.9$\,MPa produced by our model feasibly fall within the expected range. In fact, these values are approximately one order of magnitude higher than those of the effective bulk Young's modulus of the worm predicted in Ref.~\cite{Backholm_2013}, being at the same time one order of magnitude lower than those obtained from local measurements of the stiffness of the cuticle \cite{Nakajima09, Gil15}. This is consistent with the point of view that the effective mechanical properties of the worm are not exclusively defined by the stiffness of the cuticle. However, we also conclude that the cuticle plays a considerable role in the response of the worm to vibration.

\section{Conclusions}

We have demonstrated the excitation of subharmonic Faraday-like waves in living earthworms lying horizontally on a flat solid surface subjected to vertical vibration. We tested four common species of earthworm -- \textit{Eisenia fetida}, \textit{Lumbricus terrestris}, \textit{Lumbricus rubellus} and \textit{Aporrectodea caliginosa}, and in all tests we observed the appearance of spectral peaks at subharmonic frequencies with overall behaviour similar to that of finite-size liquid drops subjected to vibration. We also used an earthworm-mimicking phantom made of a water-filled cylinder with thin elastic walls, the measurements of which qualitatively reproduced the response of the real earthworms. Moreover, we measured the critical amplitude of the onset of subharmonic waves in the worms and found that it exhibits oscillations as a function of the driving frequency. This feature is typically observed in the response of infinitely extended viscoelastic fluids \cite{Bal05} and isolated small drops composed of a Newtonian fluid \cite{Yoshiyasu96,Noblin09,Ma17,ma_burton_2018}. By modelling the body of the worm as an elastic cylindrical shell filled with fluid, we explained the observed subharmonic response by parametric excitation of the discrete set of vibrational modes. We therefore conclude that the nonmonotonic dependence of the critical amplitude on the vibration frequency should be a direct consequence of the discrete nature of the spectrum of eigenfrequencies.

Because biological cells and many living organisms are mostly made of fluids, unique properties of nonlinear waves observed in fluidic systems are likely to open up unique opportunities for biology and medicine as well as the adjacent areas. The work in this direction is already in progress \cite{hong2019wavecontrolled}. Thus, we believe that our results would not only push the frontiers of our knowledge of fundamental nonlinear phenomena and chaotic behaviour in biological systems, but they could also be used to develop new techniques for probing and controlling biophysical processes inside a living body. 
\\
\begin{acknowledgments}
This work was supported by the Australian Research Council (ARC) through its Future Fellowship (FT180100343). 
\end{acknowledgments}


%

\end{document}